\documentclass[12pt,preprint]{aastex}

\shorttitle{}

\shortauthors{Yang et al.}

\begin{document}

\title{Impact of pickup ions on the shock front nonstationarity and energy dissipation of the heliospheric termination shock: Two-dimensional full particle simulations and comparison with Voyager 2 observations}

\author{Zhongwei Yang\altaffilmark{1}, Ying D. Liu\altaffilmark{1}, John D. Richardson\altaffilmark{2},
Quanming Lu\altaffilmark{3}, Can Huang\altaffilmark{3}, and Rui Wang\altaffilmark{1}}

\altaffiltext{1}{State Key Laboratory of Space Weather, National Space Science Center, Chinese Academy of Sciences, Beijing 100190, China;
liuxying@spaceweather.ac.cn}

\altaffiltext{2}{Kavli Institute for Astrophysics and Space Research, Massachusetts Institute of Technology, Cambridge, MA 02139, USA}

\altaffiltext{3}{CAS Key Laboratory of Geospace Environment, Department of Geophysics and Planetary Science, University of Science and Technology of China, Hefei, China}

\begin{abstract}

The transition between the supersonic solar wind and the subsonic heliosheath, the termination shock (TS), was observed by Voyager 2 (V2) on 2007 August 31-September 1 at a distance of 84 AU from the Sun. The data reveal multiple crossings of a complex, quasi-perpendicular supercritical shock. These experimental data are a starting point for a more sophisticated analysis that includes computer modeling of a shock in the presence of pickup ions (PUIs). Here, for the first time we present two-dimensional (2-D) particle-in-cell (PIC) simulations of the TS including PUIs self-consistently. We also report the ion velocity distribution across the TS using the Faraday cup data from V2. A relatively complete plasma and magnetic field data set from V2 gives us the opportunity to do a full comparison between the experimental data and PIC simulation results. Our results show that: (1) The nonstationarity of the shock front is mainly caused by the ripples and it can exists well even at a high percentage of PUIs (PUI\%). The width of the ripples decreases with the increasing PUI\% because the enhanced PUI foot makes the  gyro-radius of reflected solar wind ions (SWIs) become smaller. (2) PUIs play a key role in the energy dissipation of the TS, and most of the incident ion dynamic energy is transferred to the thermal energy of PUIs. (3) The PIC simulation indicates for the upstream parameters chosen for V2 conditions that the density of PUIs is about 25\% and the PUIs gain the larger fraction (approximately 86.6\%) on average of the downstream thermal pressure, consistent with V2 observations near the shock. The downstream pressure of PUIs can very in a range from about 61.9\% to 96.3\% in the direction perpendicular to the shock normal due to the impact of shock front ripples (i.e., a 2-D effect). (4) The simulated composite heliosheath ion distribution function is a superposition of a cold core formed by transmitted SWIs, the shoulders contributed by the hot reflected SWIs and directly transmitted PUIs, and the wings of the distribution dominated by the very hot reflected PUIs. It is similar to a previous theoretical heliosheath ion distribution function. (5) The V2 Faraday cups observed the cool core of the distribution, so we see only a tip of the iceberg. For the evolution of the cool core distribution function across the TS, the computed results agree reasonably well with the V2 experimental results. The relevance of the shock front ripples to the multiple crossings of TS observed by V2 is also discussed in this paper.

\end{abstract}


\section{Introduction}

The solar wind blows outward from the Sun and forms a bubble of solar material in the interstellar medium. Because the interstellar plasma confines the solar wind, it has to become subsonic before directly interacting with the interstellar plasma, and this transition occurs at the heliospheric termination shock (TS) \citep{Decker2005,Richardson2008,Jokipii2008,Burlaga2008}. Interstellar neutrals enter the heliosphere and are ionized by charge exchange with the solar wind ions (SWIs) \citep{Vasyliunas1976,Mobius1985,Lee1987,Galeev1988}. They are then picked up by the solar wind and are named pickup ions (PUIs). After these PUIs are convected to the TS, they are energized and a fraction of them become the anomalous cosmic rays (ACRs) due to diffusive shock acceleration \citep{Axford1977,Bell1978,Blandford1978,Giacalone2005,Zank2006,Fisk2006,Guo2010}. The TS is strongly influenced by the presence of PUIs \citep{Liewer1993,Gloeckler1998,Matsukiyo2007}, which makes the TS very different from planetary and interplanetary shocks inside the heliosphere \citep{Lembege2004,Burgess2005,Goncharov2014}. Of particular interest are the microstructure and the energy dissipation of the TS.

First, the number density of PUIs at the TS is relatively high ($\sim25\%$) \citep{Richardson2008,Wu2009} and thus may greatly modify the microstructure of the shock front. The TS was expected to be a boundary that is stable on a timescale of several days. The Voyager 2 (V2) plasma experiment observed a decrease in the solar wind speed commencing on about 2007 June 9, which culminated in several observations of TS crossings (named TS-2, TS-3, and TS-4) between 2007 August 30 and September 1 \citep{Burlaga2008,Decker2008,Richardson2008,Stone2008}. At least two TS crossings (TS-1 and TS-5) occurred when there were gaps in the telemetry. Observations of the magnetic field structure and dynamics of the TS were reported \citep{Burlaga2008}. The TS location depends on the solar activity, and it moves inward and outward due to changes in the solar wind pressure during the rising or declining phase of the solar cycle. This motion is on a long time scale (several years; \citealp{Whang2000}). The underlying mechanisms of how the V2 spacecraft made so many crossings in such a short time (several hours), however, still remain unclear. It is believed that the multiple crossings imply motions of the TS, which would be caused by the shock front nonstationarity \citep{Burlaga2008}.

There are many types of nonstationarity for a shock front, e.g., self-reformation \citep{Lembege1987,Hada2003,Chapman2005,Matsukiyo2014}, self-exited ripples \citep{Winske1988,Savoini1994,Lembege2004,Burgess2007}, pre-existing waves or turbulence \citep{Giacalone2005,Guo2010}, which likely cause the unexpected motions of the TS. These nonstationarities are predicted and observed by numerical simulations and satellite observations, respectively. The term ``self-reformation" describes a process that the particles reflected by the shock ramp accumulate ahead of the shock and form a shock foot, which then grows and becomes the new ramp. The new ramp starts to reflect incident particles, and the process repeats. Self-reformation of the shock front was predicted by both hybrid simulations \citep{Hellinger2002,Lembege2009,Yuan2009,Tiu2011,Su2012} and PIC simulations \citep{Lembege1987,Hada2003,Scholer2003,Nishimura2003,Lee2005a,Yang2009,Yang2013} for large Mach number and low $\beta_i$ shocks, where $\beta_i$ is the ratio of the thermal pressure of ions to the magnetic pressure. Even in the presence of PUIs, the self-reformation can still be retrieved in some conditions \citep{Lee2005b,Chapman2005,Oka2011,Yang2012a,Matsukiyo2014}. For the heliospheric TS, the values of $\beta_i$ and the Mach number are relatively low and high, respectively. The TS is generally believed to be in the supercritical regime and probably undergoes self-reformation \citep{Burlaga2008}. Self-excited ripples are usually found in 2-D hybrid simulations \citep{Thomas1989,Burgess2007} and PIC simulations \citep{Savoini1994,Lembege2009,Yang2012b}. Shock front ripples are robust in three dimensional hybrid simulations \citep{Thomas1989,Hellinger1996}. The filament instability of the shock front found in high dimensional simulations can also contribute to the shock front nonstationarity \citep{Spitkovsky2005,Guo2013,Caprioli2014}. All of the simulations above do not include PUIs. By using the 2-D Los Alamos hybrid simulation code with PUIs, \citet{Liu2010} studied the Alfv\'en-cyclotron and mirror modes excited in the near-TS heliosheath. The impact of the PUIs on the shock front ripples and self-reformation is not mentioned. It is expected that the density of PUIs at the TS is of the order of 20\% to 30\% of the solar wind density \citep{Richardson2008,Wu2009,Matsukiyo2014}. The impact of such high percentage of PUIs on the shock front ripples has not been studied yet. The relevance of the shock front ripples to the multiple TS crossings still remains unclear. A precise description of this influence would require at least a 2-D full particle model.

Second, the relative high percentage of PUIs may greatly change the dissipation mechanism of the shock front. The V2 has a working plasma instrument and the coverage was sufficient to identify three crossings of the TS. The identified TS-3 crossing revealed an almost classical perpendicular shock structure \citep{Burlaga2008,Richardson2008}. It is clearly evidenced that the TS heats the incident SWIs very little, and the average downstream ion temperature is much smaller than predicted by the MHD Rankine-Hugoniot conditions. \citet{Richardson2008} concluded that most of the solar wind energy is transferred to the PUIs or other energetic particles that reside in the energy range not covered by V2 plasma instrument. \citet{Zank1996} predicted that the TS dissipation mechanism would favor PUIs and leave the SWIs relatively cool. They concluded that PUIs may therefore provide the primary dissipation mechanism for a perpendicular TS with SWIs playing a secondary role. Previous 1-D hybrid simulations with multiple ion species \citep{Wu2009} suggest that the density of PUIs is about $25\%$ and the PUIs gain the lager fraction (approximately 90\%) of the downstream thermal pressure. They defined a downstream pressure ratio $\chi=P_t^{PUI}/P_t$, where $P_t^{PUI}$ and $P_t$ represent the thermal pressure of PUIs and the total ions, respectively. However, it is not clear how the upstream plasma dynamic energy transfers to the PUIs inside the shock front and how the downstream pressure ratio $\chi$ varies in different locations along the shock surface (i.e., a 2-D effect). In order to understand the energy dissipation process and the energy partition between different species of ions, here we develop a 2-D full particle model of the TS.

It is also worth studying how the velocity distribution function downstream of the TS looks like from PIC simulations that include PUIs. The Voyager spacecraft are making in situ measurements along two different trajectories in the heliosheath, but unfortunately they were not designed to measure the PUIs directly. Launched on 2008 October 19, IBEX \citep{McComas2009} is measuring the energetic neutral atom (ENA) flux from the boundary region of the heliosphere. The interpretation of ENA fluxes measured at 1 AU by IBEX needs knowledge of the ion velocity distribution function in the inner heliosheath. ENAs are created by charge exchange of interstellar neutrals with hot heliosheath protons or ions. The flux of ENAs therefore depends sensitively on the number of hot protons downstream of the TS. \citet{Heerikhuisen2008} find that the ENA flux for a $\kappa-$distribution is higher than that for a Maxwellian proton distribution with the same temperature. This is not surprising because the $\kappa-$distribution contains many more particles in the wings of the distribution function than the corresponding Maxwellian distribution. Why the heliosheath proton distribution function should be like a $\kappa-$distribution is, however, unclear. The answer may well reside in the processing of upstream PUI distribution by the TS and subsequent relaxation of the processed distribution in the heliosheath.

In this paper, we use a 2-D PIC code to investigate the impact of PUIs on the shock front microstructure, the energy dissipation, and the downstream particle velocity distribution function of the TS. This paper is organized as follows. The simulation model is described in Section 2. We describe the simulation results in Section 3, focusing on the impact of PUIs on the shock front nonstationarity and the particle energy partition. In Section 4, we compare the simulation results with V2 observations. Finally, conclusions are drawn in Section 5.

\section{Simulation model}

We use a 2-D electromagnetic PIC code to simulate the evolving structures of supercritical, collisionless, perpendicular shocks with PUIs. Simulations of nonstationary shocks have been performed by 1-D PIC codes where PUIs are included \citep{Chapman2005,Matsukiyo2011,Yang2012a}. In this paper, we expand our simulation code to 2-D in order to investigate the ripples of the shock front. Here, the control equations of the PIC code are Maxwell and Newton-Lorentz equations only. Due to the numerical error built up during PIC simulations, Guass¡¯ law cannot be assumed to be satisfied all the time. Instead of solving Possion equation, our code solves only two curls, i.e., Ampere and Faraday equations. A rigorous charge conservation method for the current deposits is described in \citet{Villasenor1992}. By using the charge conservation method, an additional equation $\partial{\rho}/\partial{t}=-\nabla\cdot \vec{J}$ should be solved at each time step to provide the current density to the field update. This method has been commonly used in previous PIC simulations \citep{Buneman1993,Nishikawa1997,Cai2003,Spitkovsky2008,Rekaa2014}. The particle data are updated by the leap-frog method \citep{Birdsall1991}.

For the 2-D simulations, we consider a Cartesian grid ($x$,$y$). The plasma box size along the shock normal and shock front is $L_x=4096\Delta_x$ and $L_y=512\Delta_y$, respectively, where the numerical grid spacing $\Delta_x=\Delta_y=0.025c/\omega_{pi}=0.25c/\omega_{pe}$. The spatial resolution is high enough to resolve the microstructures of the shock front even on the electron inertial scale \citep{Mazelle2010}. The physical vectors, such as velocities of particles, and electric and magnetic fields $E$ and $B$, have components in three directions and also spatially depend on $x$ and $y$. A shock is produced by using the so-called piston method \citep{Burgess2007,Hao2014}, in which the plasma is injected continuously from one end of the simulation box ($x=0$, in our case), and reflected elastically at the other end ($x=L_x$). The upstream plasma has a uniform density with 16 particles per species per cell. The fractions of the particles of different species can be changed for different purposes (e.g., for a 25\% PUI case, the fractions of electrons, SWIs, and PUIs are 1, 0.75 and 0.25 respectively). The right boundary is assumed to be a perfectly conducting barrier. The pileup of the density and magnetic field creates a shock propagating in the $-x$ direction. In the 2-D simulation, the boundary conditions of the electromagnetic fields in the $y$ direction are periodic, and the particles that move out of one end of the simulation domain in the $y$ direction will re-enter the domain from the other end. The initial distribution functions for the SWIs and electrons are both Maxwellian. PUIs are injected on a thin sphere in velocity space centered at $V_{inj}$ with a radius $V_{shell}$ as in earlier 1-D works \citep{Chapman2005,Yang2012a,Matsukiyo2014}. The upstream Alfv\'en speed $V_A$ is equal to 1. To reduce the computational time, we use an unrealistic mass ratio for ions and electrons $m_i/m_e=100$, and light speed $c=20V_A$ as usual \citep{Chapman2005,Oka2011}. The basic parameters and configuration are as follows. The ambient magnetic field $B_o$ is in the $Y$ direction as in previous work \citep{Winske1988,Liu2010}. The solar wind ion $\beta_i$ is 0.04 as observed by Voyager 2 \citep{Burlaga2008,Richardson2008}. The injected plasma is quasi-neutral, i.e., $n_i=n_e$ and $n_i=n_{SWI}+n_{PUI}$, where $n_e$, $n_i$, $n_{SWI}$, and $n_{PUI}$ are densities of the electrons, total ions, SWIs and PUIs, respectively.

\section{Simulation results on the impact of PUIs}

In this section, the impact of PUIs on the shock front nonstationarity will be analyzed in detail for three cases with different percentages of PUIs (PUI\%): 0\% (Run A), 10\% (Run B), and 25\% (Run C). Then we concentrate on an open question of how the upstream ion dynamic energy transfers to the thermal energy of PUIs and SWIs and the magnetic energy within the shock transition layer. Furthermore, the energy partition of SWIs and PUIs downstream of the TS will also be computed as in previous 1-D hybrid simulations \citep{Wu2009}.

\begin{figure}
\epsscale{0.6} \plotone{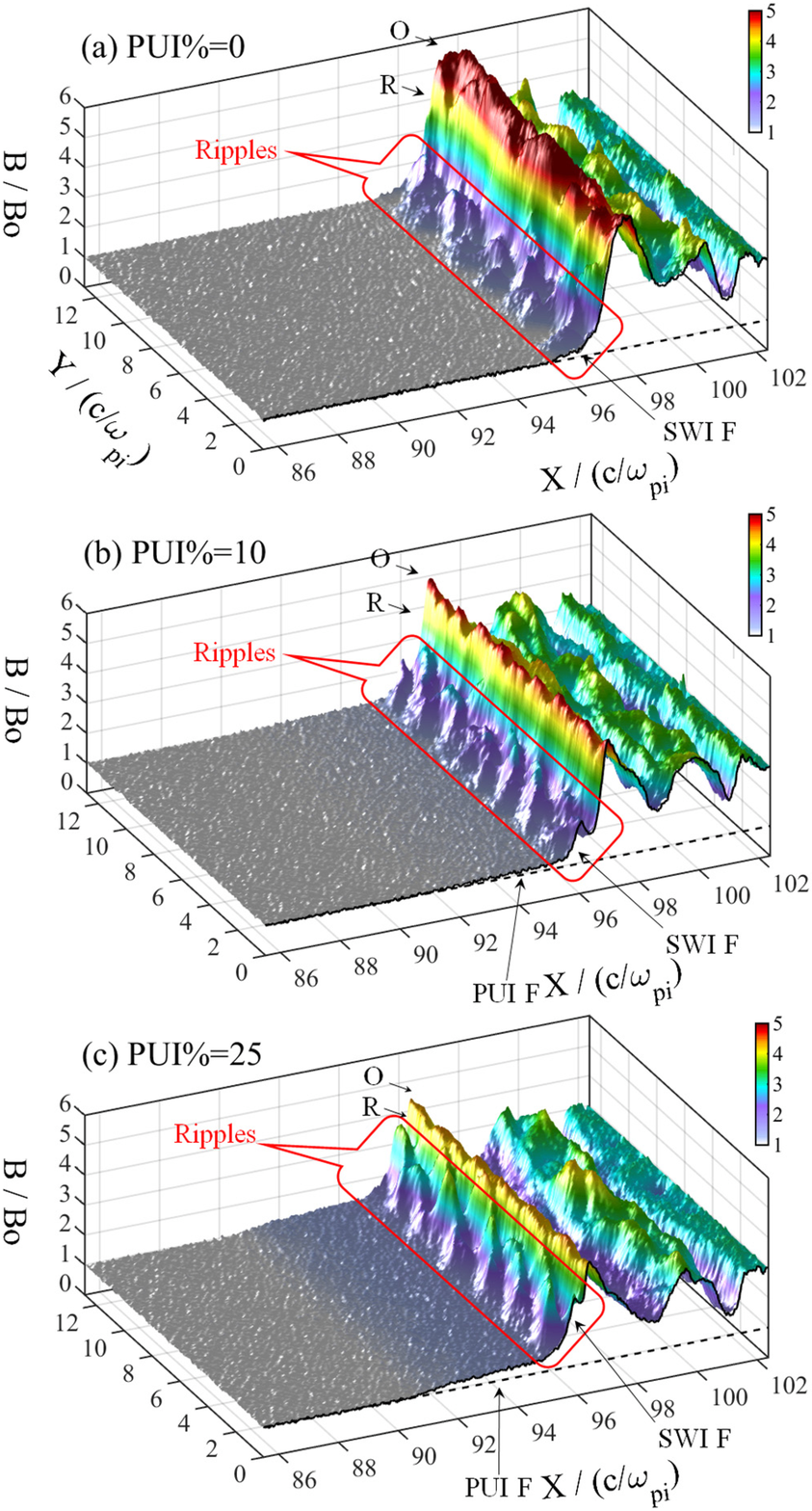}
\caption{\label{fig1}Overview of the magnetic field profile at a time $t=4\ \Omega_{ci}^{-1}$ resulting from different Runs: (a) PUI\%=0, (b) PUI\%=10, and (c) PUI\%=25. The color shading indicates the strength of the magnetic field. The shock foot, ramp and overshoot are marked by ``F", ``R" and ``O", respectively. The foot regions due to the reflected PUIs and SWIs are labeled as ``PUI F" and ``SWI F", respectively. The dashed black reference line shows the upstream value $B_o$. The $B$ profiles at $Y=0$ are shown by black solid curves. The shock front ripples are marked by a red box.}
\end{figure}

First, we investigate the impact of the relative percentage of PUIs on the shock front. Figure \ref{fig1} is an overview of Runs A (PUI\%=0), B (PUI\%=10), and C (PUI\%=25). In each panel, the surface indicates the magnetic field $B$ in the 2-D simulation domain at a time $t=4\ \Omega_{ci}^{-1}$. Figure \ref{fig1}a shows the shock case in the absence of PUIs. In this case, the shock front is characterized by self-excited ripples (marked by a red box) as in previous 2-D PIC simulations without PUIs \citep{Savoini1994,Lembege2009}. Typical structures of a supercritical perpendicular shock such as the foot (``F"), ramp (``R") and overshoot (``O") are evident in this case (see the magnetic field profile at $Y=0$ marked by the black solid curve). The upstream Alfv\'enic Mach number $M_A$ is about 4.5. A similar plot for Run B is shown in Figure \ref{fig1}b. In this case, a weak but broad PUI foot with an average amplitude of $\sim1.15\ B_o$ (marked by ``PUI F") emerges ahead of the SWI foot (marked by ``SWI F"). The PUI foot is stationary. Figure \ref{fig1}c shows the corresponding plot for Run C. In this high PUI\% (=25) case, the amplitude of the PUI foot becomes higher, but the amplitude of the overshoot becomes lower than that in Runs A and B.

\begin{figure}
\epsscale{0.65} \plotone{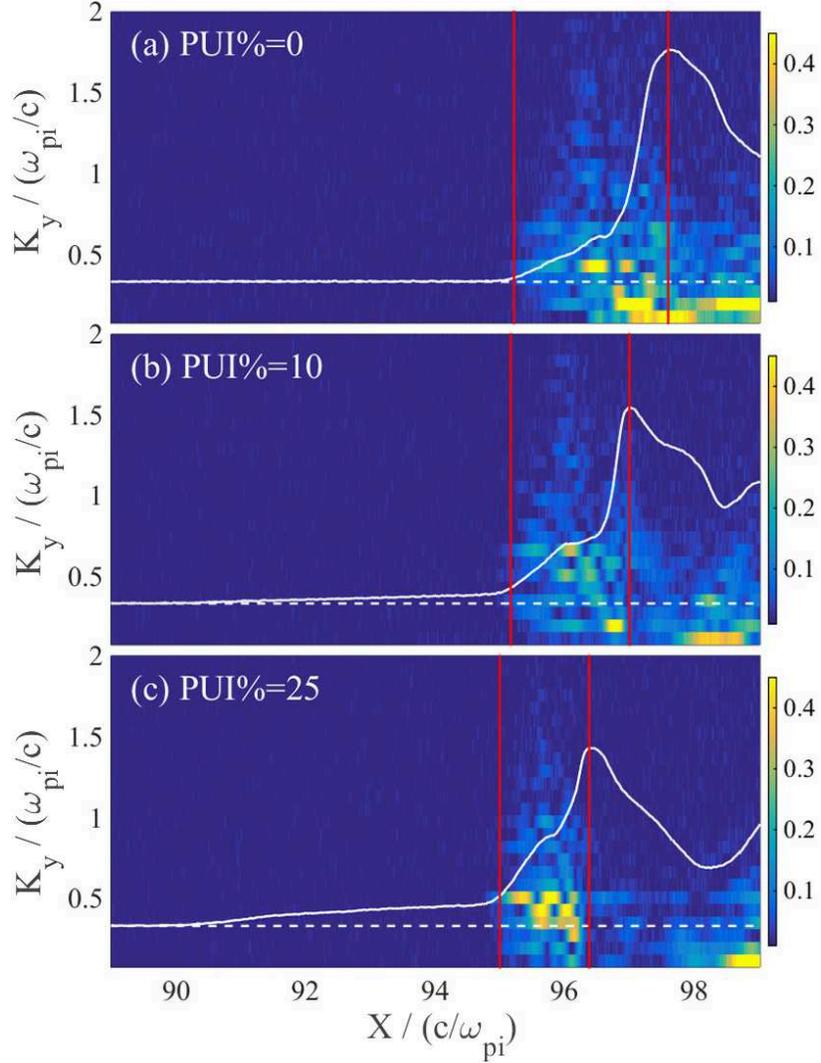}
\caption{\label{fig2}The power spectrum of the fluctuating magnetic field $B$ across the shock with 0\%, 10\% and 25\% PUIs (from top to bottom) at $t=4\ \Omega_{ci}^{-1}$. The color shading indicates the power ($|B(K_y)|^2$) which is obtained by Fourier transforming the values of $B$ along the $Y$ direction at a selected position $X$. The vertical red lines denote the interval of the rippling region. The profile $\bar{B}/3$ (averaged along the $Y$ direction) is superimposed.}
\end{figure}

As can be seen from Figure \ref{fig1}, the shock front is nonstationary even in the high PUI\% case due to the self-excited ripples. The impact of the relative percentage of PUIs on the rippling shock front can be illustrated from two aspects: wave features and particle behaviors. Figure \ref{fig2} plots the corresponding power spectrum of the fluctuating magnetic field $B$. The color shading indicates the power $|B(k_y)|^2$ which is obtained by Fourier transforming the values of $B$ along the $Y$ direction (i.e., along the shock front) at a selected position $X$. In order to show the location of the shock front, the $Y$-averaged magnetic field $B/3$ has been overlaid on the spectrum for reference (white curve). The horizonal dashed line indicates the upstream value of $B/3$. The magnetic field fluctuations are enhanced greatly form the SWI foot to the ramp. With the increase of PUI\%, the ripple excitation region in the $x$ direction (between the two vertical red lines) becomes narrower.

\begin{figure}
\epsscale{0.65} \plotone{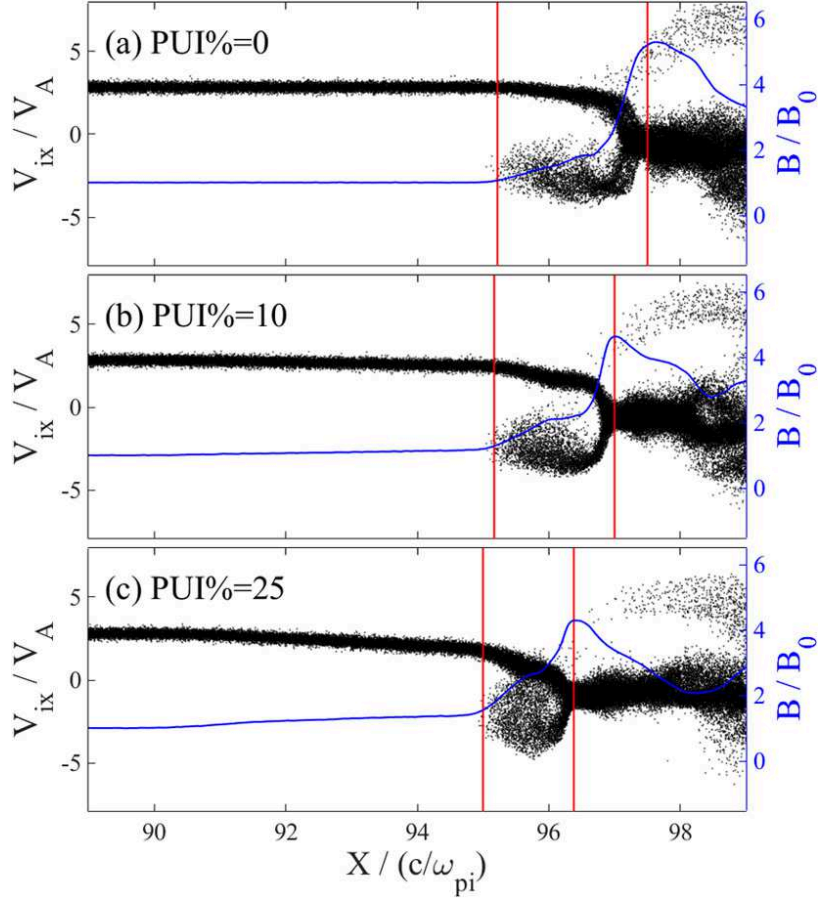}
\caption{\label{fig3} Phase space plots ($V_{ix}$ versus $X$) of SWIs at $t=4\ \Omega_{pi}^{-1}$. From top to bottom, the relative density of PUIs is 0\%, 10\%, and  25\%, respectively. In each panel, the Y-averaged $B$ field (blue curve) is shown for reference. The ripple region is between the two vertical red lines.}
\end{figure}

To understand why the scale of ripples along $Y$ decreases with the relative percentage of PUIs, we perform particle diagnosis for the cases in Figure \ref{fig2}. Corresponding phase space diagrams ($X-V_{ix}$, black dots) of SWIs are plotted in the downstream rest frame (i.e., the simulation frame) in Figure \ref{fig3}. The $Y-$averaged magnetic field $B$ is also shown for reference (blue curve). A fraction of the incident SWIs coming from the left-hand side is reflected at the shock front, and the reflected ions become a hot SWI population when convected back to the downstream. The other incident SWIs are directly transmitted to the downstream region and form the cool core of the downstream ion velocity distribution. The shock front ripple is associated with the reflected SWIs. In brief, the magnetic field of the PUI foot ahead of the shock ramp increases with the percentage of PUIs. In a high PUI\% case, the gyro-radius of the R SWIs becomes smaller due to the enhanced local $B$ caused by the R PUIs at the foot. This leads to a narrower ripple excitation region at the shock front.

\begin{figure}
\epsscale{0.75} \plotone{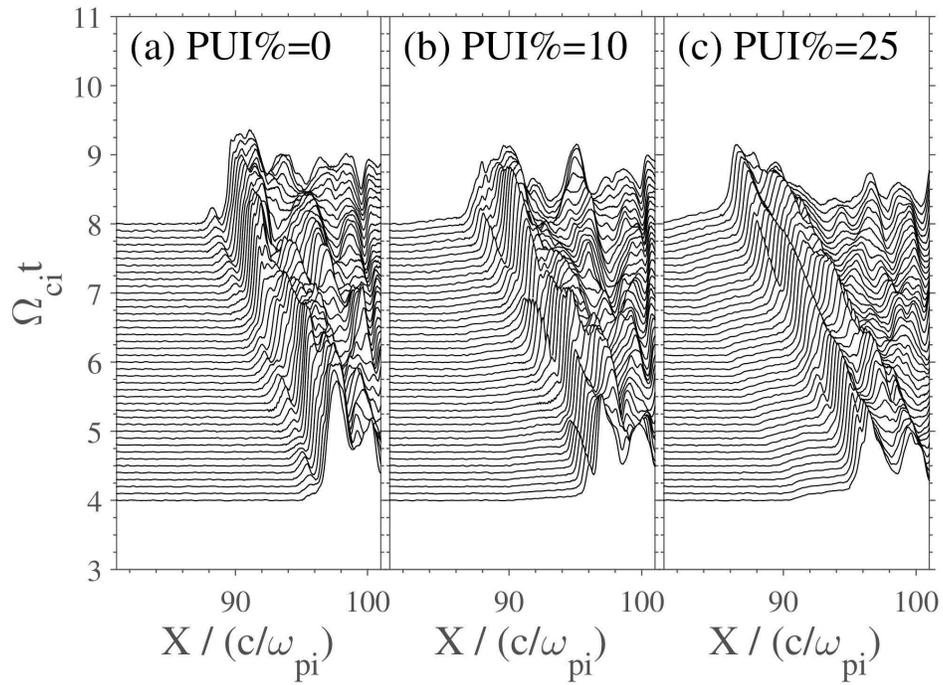}
\caption{\label{fig4}Stack-plots of the $B$ profiles (at $Y=0$) at different times. From left to right, the relative density of PUIs is 0\%, 10\%, and 25\%, respectively.}
\end{figure}

Figure \ref{fig4} shows stack plots of the $B$ profiles at a fixed $Y$ ($=0$) versus time with different PUI\%. The period of the shock front nonstationarity is about $1-2\ \Omega_{ci}^{-1}$, which is close to the self-reformation time observed in 1-D PIC simulations \citep{Hada2003,Chapman2005,Yang2009}. The time scale of the first TS crossing (named TS-2) observed by V2 is about 29 min \citep{Richardson2008}, which is equivalent to $1.7\ \Omega_{ci}^{-1}$. Our PIC simulation give a similar period, so TS-2 is likely caused by the self-excited rippling at the shock front. The time intervals of other two crossings (TS-3 and TS-4) are about 3.7 hours and 2.8 hours ($\sim10\ \Omega_{ci}^{-1}$), respectively. These crossings are likely due to the interaction of the TS with the pre-existing waves or turbulence with a larger temporal scale \citep{Guo2012}.

Second, we study the energy dissipation process within the shock transition layer and the resulting energy partition in the downstream region of the TS. \citet{Wu2009} studied the energy partition of PUIs and SWIs at the TS by 1-D hybrid simulations. For comparison with V2 observations, they defined the energy partition of PUIs in the downstream as a pressure ratio $\chi=P_t^{PUI}/P_{t}$, where $P_t$ and $P_t^{PUI}$ are the downstream thermal pressure of the total ions and PUIs, respectively. The pressure ratio $\chi$ increases as the PUI relative density increases. In the high percentage (25\%) PUI case, the ratio $\chi$ is about 90\%, which is the energy fraction gain for PUIs inferred from the Voyager 2 observations by \citet{Richardson2008}. However, 1-D hybrid simulations have the fundamental limitation that the downstream heating occurs only in the two directions perpendicular to the magnetic field \citep{Wu2009}. 2-D hybrid simulations of quasi-perpendicular shocks without PUIs have demonstrated downstream ion temperature anisotropies and associated cyclotron and mirror-like fluctuations \citep{Winske1988,Lu2006,Hao2014}, consistent with observations \citep{Liu2006,Liu2007,Richardson2007}. Here, we examine the energy partition of PUIs in the downstream of the TS by using a 2-D PIC code which allows the heating to take place in all three directions.

\begin{figure}
\epsscale{0.7} \plotone{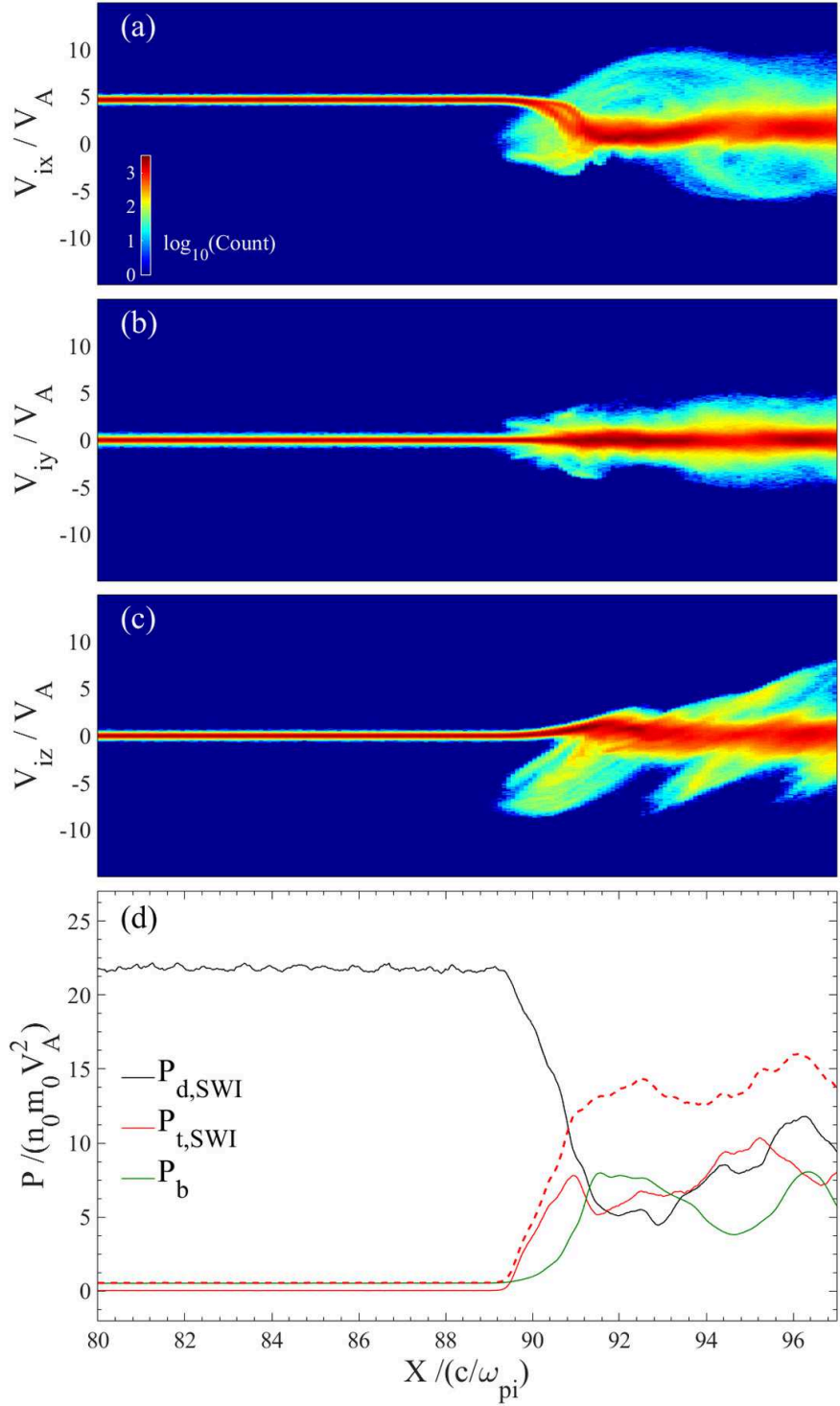}
\caption{\label{fig5}(a) Phase space plots ($X-V_{ix}$) of the SWIs at $t=7\ \Omega_{ci}^{-1}$ in Run A (PUI\%=0). The color shading shows logarithmic distributions
of the particle number. (b-c) Similar plots but for $X-V_{iy}$ and $X-V_{iz}$, respectively. (d) The dynamic pressure of SWIs (black), the thermal pressure of SWIs (red), and the magnetic pressure (green) across the shock front. The sum of the SWI thermal pressure and the magnetic pressure is also shown for reference (red dashed).}
\end{figure}

For reference, we present the energy dissipation at a shock without PUIs. Figure \ref{fig5} shows the phase space plots ($X-V_{ix}$, $V_{iy}$ and $V_{iz}$) of SWIs in the shock rest frame for Run A. Note that the simulations are performed in the downstream rest frame so that the downstream flow speed $|V_{DS}|=|V_{s}|$, where $V_s$ is the shock propagation speed. The top three panels in Figure \ref{fig5} show that the SWIs are mainly heated along the directions perpendicular to the magnetic field. The dynamic pressure of the SWIs along the shock normal can be calculated from $P_{d}=nmV_x^2$, where $n$, $m$, and $V_x$ are the density, mass, and bulk velocity of ions in the $X$ direction. The thermal pressure of SWIs along the shock normal can be calculated from $P_{t}=nkT$, where $k$ and $T$ are the Boltzmann constant and the temperature of the particles. The magnetic pressure $P_b$ is computed from $B^2/(2\mu_0)$. All the pressures are normalized by $n_0m_0V_A^2$ of the upstream region (Figure \ref{fig5}d), where $m_0$ and $n_0$ are the mass and upstream density of the ions, respectively. A reduction of the dynamic pressure of SWIs is evident at the shock transition. Both of $P_{t}$ and $P_b$ increase at the the shock transition, and their sum is negatively correlated with the dynamic pressure $P_d$. Obviously, the upstream dynamic energy is transferred to the SWI thermal energy and the magnetic energy of in the shock transition.

\begin{figure}
\epsscale{0.7} \plotone{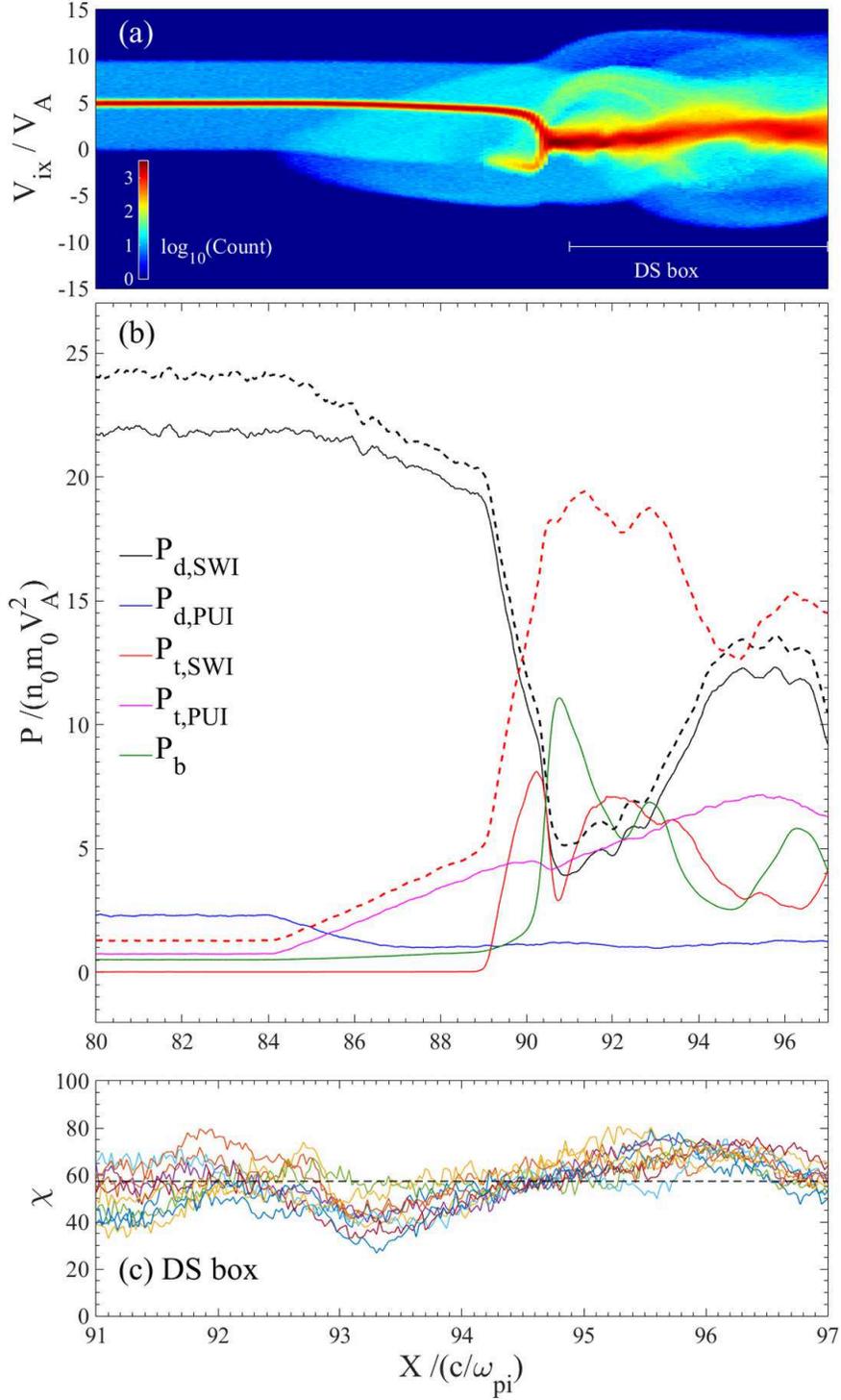}
\caption{\label{fig6}(a) Phase space plots of the SWIs and PUIs at $t=7\ \Omega_{ci}^{-1}$ in Run B (PUI\%=10). The color shading shows logarithmic distributions of the particle number. (b) The dynamic pressure of SWIs (black) and PUIs (blue), the thermal pressure of SWIs (red) and PUIs (magenta), and the magnetic pressure (green) across the shock front. The black dashed curve indicates the total ion dynamic pressure. The red dashed curve indicates the sum of the total ion thermal pressure and the magnetic pressure. (c) The pressure ratio $\chi=P_t^{PUI}/P_t$ computed in a downstream region (marked by ``DS box" in the top panel).}
\end{figure}

The impact of PUIs on the dissipation process and the energy partition is shown in Figure \ref{fig6} with a moderate PUI\% (10). In the PUI foot region (from $X=84\ c/\omega_{pi}$ to about $X=89\ c/\omega_{pi}$), the dynamic pressures of both PUIs and SWIs start to decrease. Simultaneously, the thermal pressure of PUIs increases due to the reflected PUIs (Figure \ref{fig6}a). There are no reflected SWIs in the PUI foot (see Figure \ref{fig6}a), thus the thermal pressure of SWIs is almost unchanged in this region. The magnetic pressure $P_b$ slightly increases at the PUI foot. Therefore, most of the decreased dynamic energy of the incident ions (PUIs plus SWIs) is transferred to the PUIs and dissipated as their thermal energy in the PUI foot. In the SWI foot region (from $X=89\ c/\omega_{pi}$ to about $X=90.25\ c/\omega_{pi}$), the dynamic pressure of SWIs decreases drastically due to the reflected SWIs, while the dynamic pressure of PUIs is almost unchanged. The thermal pressure of PUIs and the magnetic field pressure increase gradually. Thus, most of the energy of SWIs is dissipated as the thermal energy of SWIs in the SWI foot. In the narrow ramp region (from $X=90.25\ c/\omega_{pi}$ to the about $X=90.75\ c/\omega_{pi}$), all the pressures decrease except the magnetic pressure, which shows a steep increase. Thus, the dynamic energy and the thermal energy of the total ions are transferred to the magnetic field in this region. Figure \ref{fig6}c shows the thermal pressure ratio $\chi$ in a downstream region. Curves in different colors indicate the $\chi$ profiles obtained in different $Y$ locations. The ratio $\chi$ can vary versus $Y$ due to the downstream turbulence as a remnant of the shock front ripples (a 2-D effect). This can not be obtained in 1-D simulations because of $\partial/\partial Y=0$. The average value of the ratio $\chi$ in the current 2-D simulations is about 57.3\% (marked by a horizonal black dashed line).

\begin{figure}
\epsscale{0.7} \plotone{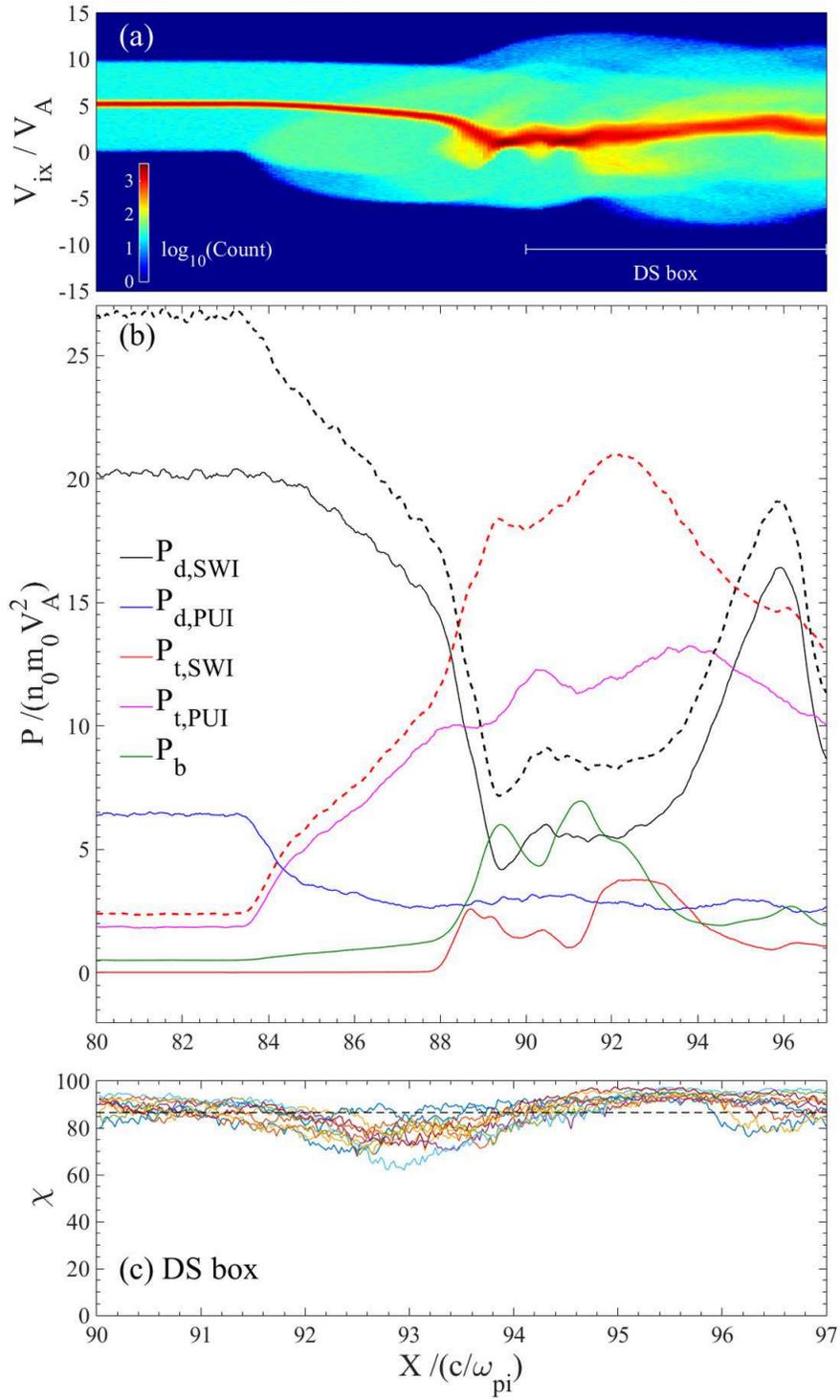}
\caption{\label{fig7} The same format as Figure 6, but for Run C (PUI\%=25).}
\end{figure}

Then we do similar calculations for the high percentage PUI (25\%) case, as shown in Figure \ref{fig7}. In contrast to the low percentage PUI (10\%) case, the SWIs lose almost half of their dynamic energy at the PUI foot (from $X=83.75\ c/\omega_{pi}$ to about $X=88\ c/\omega_{pi}$). As a consequence, the PUIs gain more thermal energy in this region. The gains of SWI thermal energy and the magnetic energy are lower than those in Run B. Most of the upstream dynamic energy is transferred to PUIs, and the pressure of PUIs becomes much higher than that of SWIs. In this case, the ratio $\chi$ can vary in a range from about 61.9\% to 96.3\% along the $Y$ direction due to the remnant effect of shock front ripples. The average value of the pressure ratio $\chi$ in the downstream region is about 86.6\%, which is consistent with previous 1-D hybrid models \citep{Wu2009} and the V2 experimental data \citep{Richardson2008}.

\section{Comparison with Voyager 2 observations}

\begin{figure}
\epsscale{0.55} \plotone{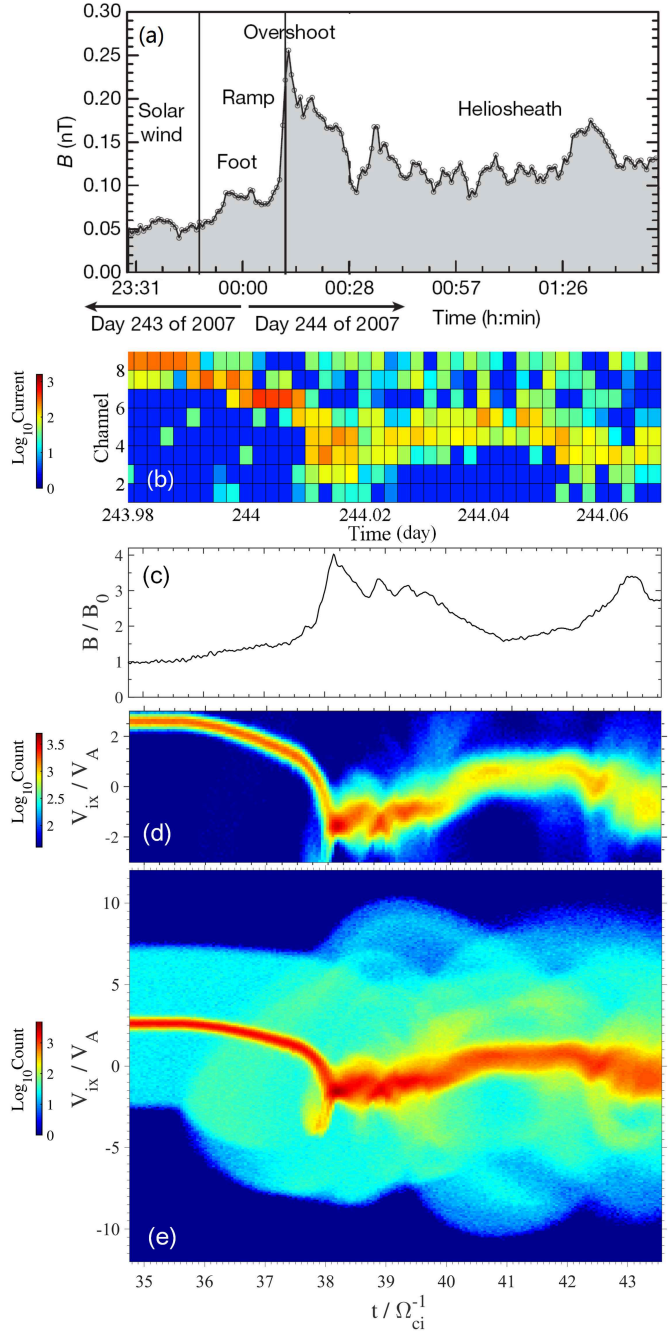}
\caption{\label{fig8} (a) 48-s averages of the magnetic field strength $B$ at the TS (TS-3, after \citealp{Burlaga2008}). (b) Velocity distribution observed by Faraday cups on V2. Different channels correspond to different velocity ranges. The color shading shows logarithmic distributions of the current or particle count number. (c) The in situ magnetic field strength $B$ seen by a virtual probe (VP) across the TS profiles obtained from Run C (PUI\%=25). (d) Corresponding logarithmic distributions of the ions seen by the VP across the TS. Here the particle counts range from 40 to 5000. (e) Similar plot as in panel (d) but with a full count sensitivity (counts range from 0 to 5000). It shows the complete distribution of the total ions across the TS.}
\end{figure}

We compare the simulation results with V2 observations. Figure \ref{fig8}a shows the observed $B$ field at the TS-3 crossing (after \citealp{Burlaga2008}). The identified TS-3 crossing revealed an almost classical perpendicular shock structure. The plasma instrument (Faraday cups) on V2 worked well during the TS crossings. Figure \ref{fig8}b shows the observed low energy ion data. The average speed corresponding to the channel numbers 1, 2, 3, 4, 5, 6, 7, 8, 9, and 10 is about 60, 90, 119, 148, 216, 256, 300, 352, and 410 km/s, respectively. The bulk speed of the low energy ions decreases at the foot and reaches a very low value after the shock front. The estimated upstream ion gyro-period is about 17 min \citep{Burlaga2008}. Thus the time span of the TS-3 crossing plotted in Figure \ref{fig8} is about $8.8\ \Omega_{ci}^{-1}$. The V2 spacecraft moved outwards with a speed $\sim18$ km/s, and the TS-3 front moved inwards with a speed $\sim68$ km/s. Hence, the relative speed between V2 and the shock is about $86$ km/s\ $\sim 2.28\ V_A$, where $V_A$ is the Alf\'ven speed measured in the upstream of the TS at a distance of 84 AU from the sun.

In order to generate time series in PIC simulations, a virtual probe (``VP") is used which records the in situ electromagnetic field and plasma information as done in previous work for Cluster crossings of the Earth's bow shock \citep{Scholer2006}. The VP moves from the upstream to the downstream of the shock in Run C (PUI\%=25). The relative speed between the VP and the shock front is about $2.2\ V_A$. The in situ $B$ field seen by the VP is shown in Figure \ref{fig8}c. The corresponding phase space plot of the ions (SWIs plus PUIs) is shown in Figure \ref{fig8}d. Because the low energy plasma instrument on V2 is not sensitive enough to see all of the ions (e.g., the wings of the ion velocity distribution), V2 can not directly observe the PUIs. To imitate this effect, the color bar range of Figure \ref{fig8}d has been set from 40 to 5000. It means that the number of ions lower than 40 in the phase space can not be seen by the VP. Figure \ref{fig8}d shows that the simulated ion distribution across the shock is quite similar to that observed by V2 (Figure \ref{fig8}b). If we set the color bar range from 0 to 5000, the VP can see all of the ions during the shock crossing (Figure \ref{fig8}e). It implies that the thermal speed of the total ions should be higher than that estimated by using V2 plasma data. To give a more clearer view, the downstream ion velocity distribution function is computed from the overshoot to the far downstream of the shock.

\begin{figure}
\epsscale{0.98} \plotone{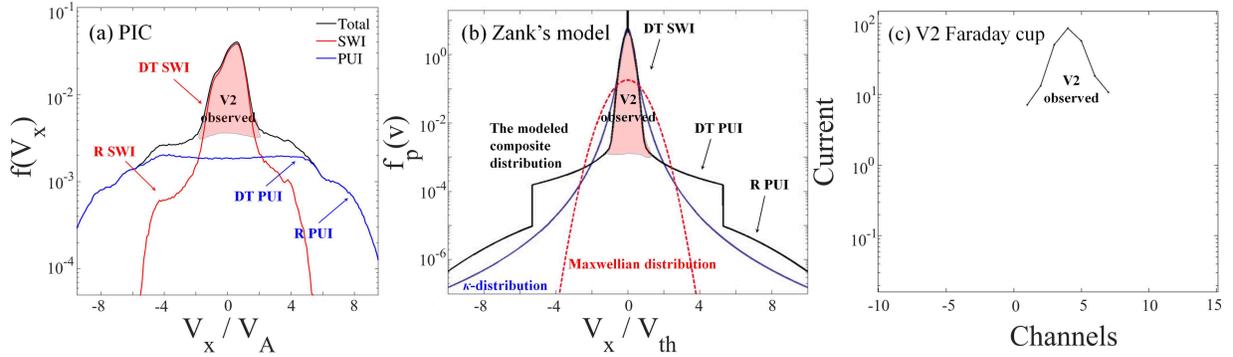}
\caption{\label{fig9} (a) Velocity distribution of the ions (black) selected in the ``DS box" (see Figure \ref{fig7}) of the TS. Contributions of SWIs and PUIs to the total distribution are indicated by red and blue curves, respectively. The highlight region (cool core) at the peak of the distribution is expected to be observed by V2. (b) Heliosheath particle distribution (black) modeled by \citet{Zank2010}. The Maxwellian (red) and $\kappa$ (blue) distributions with the same temperature and particle density are also shown for reference. (c) Cool core of the ion distribution observed by V2 in the downstream region of TS-3.}
\end{figure}

Figure \ref{fig9}a shows the normalized total ion velocity distribution (black curve) in the downstream in Run C. The total heliosheath ion distribution function is a superposition of cold transmitted SWIs (DT SWI), hot reflected SWIs (R SWI), hot directly transmitted PUIs (DT PUI), and a very hot PUI population (R PUI) that is reflected by the TS and then convected back to the downstream. The highlighted region in red is the part expected to be observed by V2. Figure \ref{fig9}b shows the composite heliosheath ion distribution function modeled by \citet{Zank2010}. Their form of the total ion distribution function is similar to that obtained from our PIC simulations (Figure \ref{fig9}a). For reference, the blue solid and red dashed curves illustrates a $\kappa-$distribution and a Maxwellian distribution with the same downstream density and temperature. Both of the modeled composite distribution and the $\kappa-$distribution have more ions in the wings of the distribution than a corresponding Maxwellian one. The R SWI population is not included in Zank's model; our simulations show that the contribution of R SWIs is small compared with that of R PUIs. Previous 1D PIC simulations have also shown that the fraction of R SWIs decreases with the PUI\% at the shock \citep{Yang2012a}. Figure \ref{fig9}c shows the ion velocity distribution observed by the Faraday cup on V2 in the heliosheath near the TS. Compared with the experimental data, both of velocity distribution functions obtained from the PIC simulations and Zank's models imply that V2 only saw the tip of the iceberg (i.e., the cool core of the total distribution). This cool part of the distribution can tell the bulk velocity of the plasma, the number density of SWIs, and roughly the temperature of SWIs.

\section{Conclusions and Discussion}

In this paper, we use a 2-D PIC code to investigate the impact of PUIs on the nonstationarity and energy dissipation of the TS. We summarize our main findings below:

1. Different from previous 1-D simulations, we have shown that the shock front ripples can still exist even when the relative percentage of pickup ions approaches 25\%. The excitation of ripples is associated with the reflected solar wind ions. In a high percentage (25\%) pickup ion case, the gyro-radius of the reflected solar wind ions becomes smaller due to the enhanced local $B$ caused by the reflected pickup ions at the shock foot. This leads to a narrower ripple excitation region at the shock front. The period of the shock front nonstationarity caused by the ripples is about $1-2\ \Omega_{ci}^{-1}$. The multiple crossings of the TS in a short time duration (e.g., TS-2) can be explained by the shock front nonstationarity.

2. The energy dissipation process is examined for shocks with different percentages of pickup ions. At a shock in the absence of pickup ions, the transformation of wave-particle energy shows that the dynamic energy at the shock front is transferred to the thermal energy of solar wind ions and the magnetic energy. In contrast, most of the dynamic energy is transferred to the thermal energy of pickup ions instead of solar wind ions at a shock in the presence of 25\% pickup ions.

3. In order to examine the energy partition in the downstream of the shock with pickup ions, we compute the pressure ratio $\chi$, as in previous 1-D hybrid simulations. In the 10\% pickup ion case, the ratio is about 57.3\%. For the 25\% pickup ion case, the average value of the ratio $\chi$ reaches about 86.6\% and can vary in a range from about 61.9\% to 96.3\% along the $Y$ direction due to the remnant effect of shock front ripples (a 2-D effect). In 1-D simulations, the ratio $\chi$ can not vary along the $Y$ direction because of $\partial/\partial Y=0$. The y-averaged ratio $\bar \chi$ obtained in the 2-D simulation is consistent with that obtained in previous 1-D hybrid simulations and that estimated by the Voyager 2 experimental data.

4. We compare our 2-D PIC simulations with the magnetic field and plasma observations for a typical shock crossing (TS-3). The velocity distribution of low energy solar wind ions resulting from PIC simulations are quit similar to the Voyager 2 observed plasma data. In addition, the velocity distribution of the pickup ions, which can not be directly observed by Voyager 2, is also predicted future observations and studies.

5. A composite heliosheath ion distribution function is obtained in our simulation. The core of the distribution function is formed by the cool directly transmitted solar wind ions, the shoulder is contributed by the hot directly transmitted pickup ions and reflected solar wind ions together, and the wing of the distribution is dominated by the very hot reflected pickup ions. The shape of the total distribution is similar to the theoretical model made by \citet{Zank2010}. Compared with the experimental data, both of the velocity distribution functions obtained from the PIC simulations and Zank's model imply that Voyager 2 only observed the tip of the iceberg (i.e., the cool core of the total distribution). The PIC simulation results may help interpret the IBEX data in probing the microphysics of the termination shock.

By using a fixed shock profile, \citet{Burrows2010} found that multiply reflected pickup ions (i.e., shock surfing accelerated pickup ions) could account for the termination shock downstream energy gains that are generally assumed to go into the pickup ions. We find that the termination shock is nonstationary and the shock front width changes with time. The shock surfing may be sufficient only when the termination shock is steep and has a narrow shock profile \citep{Zank1996,Lee1996,Lipatov1999,Shapiro2003,Yang2009}.

\acknowledgments The authors are grateful to B. Lemb\`ege and F. Guo for helpful discussions. This research was supported by NSFC under grants No. 41204106 and 41374173, the Recruitment Program of Global Experts of China, the Specialized Research Fund for State Key Laboratories of China, and the Shanghai Science Foundation (12ZR1451500).

\bibliography{mybib}
\bibliographystyle{apj}

\clearpage

\clearpage

\clearpage

\clearpage

\clearpage

\clearpage

\clearpage

\clearpage

\clearpage

\end{document}